\documentclass{appolb}
\usepackage{xcolor,graphicx,float}
\usepackage{caption,lineno}
\usepackage[colorlinks]{hyperref}
\usepackage[numbers,sort]{natbib}
\hypersetup{linkcolor=blue, citecolor=blue, filecolor=blue, urlcolor=blue}

\newcommand{\sNN}{$\sqrt{s_{\mathrm{NN}}}$}
\begin{document}
\title{Measurement of Light Nuclei Production in Heavy-ion Collisions by the STAR experiment
\thanks{Presented at XXIXth International Conference on Ultra-relativistic Nucleus-Nucleus Collisions, Krakow, Poland, April 4-10, 2022.}
}
\author{Hui Liu (for the STAR Collaboration)
\address{Key Laboratory of Quark \& Lepton Physics (MOE) and Institute of Particle Physics, Central China Normal University, Wuhan 430079, China}
\\[-4mm]
\address{and}
\\[-4mm]
\address{Physics Institute, Heidelberg University, Heidelberg 69120, Germany}
}
\maketitle
\begin{abstract}
In these proceedings, we present the measurements of centrality, transverse momentum and rapidity dependences of proton ($p$) and light nuclei ($d$ ($\overline{d}$), $t$, $^{3}\mathrm{He}$ ($\overline{^{3}\mathrm{He}}$), and $^{4}\mathrm{He}$) production in Au+Au collisions at $\sqrt{s_{\mathrm{NN}}}$ = 3 GeV, and isobaric (Ru+Ru and Zr+Zr) collisions at $\sqrt{s_{\mathrm{NN}}}$ = 200 GeV. The compound yield ratios in central collisions at 3 GeV are found to be larger than the transport model calculations. Furthermore, the kinetic freeze-out parameters at 3 GeV show a different trend compared to those of light hadrons ($\pi$, $K$, $p$) at higher energies.

\end{abstract}
  
\section{Introduction}
The Beam Energy Scan (BES) program at the Relativistic Heavy-ion Collider (RHIC) aims at understanding the phase structure and properties of strongly interacting matter under extreme conditions. In particular, it is designed to map out the first order phase transition boundary and search for the possible QCD critical point (CP) of the phase transition from hadron gas to quark-gluon plasma (QGP)~\cite{STAR:2010vob,Aoki:2006we,Luo:2017faz}.
 
Light nuclei production is predicted to be sensitive to the QCD phase structure in heavy-ion collisions~\cite{STAR:2011eej,Sun:2018jhg,ALICE:2015wav}. The STAR experiment has measured the production of deuteron~\cite{STAR:2019sjh} and triton~\cite{Zhang:2020ewj} in Au+Au collisions at {\sNN} = 7.7, 11.5, 14.5, 19.6, 27, 39, 54.4, 62.4, and 200 GeV from the first phase of RHIC BES program. The measurements of light nuclei production presented in these proceedings are obtained from Au+Au collisions at {\sNN} = 3 GeV and isobaric (Ru+Ru and Zr+Zr) collisions at {\sNN} = 200 GeV, both of which were taken in 2018.  

\section{Analysis Details}
The particle identification at low transverse momenta ($p_{T}$) is performed via their specific energy loss measured by the Time Projection Chamber (TPC). At higher transverse momenta, the particle identification is performed using the Time of Flight (TOF) detector. The final spectra of particles are obtained by correcting for the TPC tracking efficiency, TOF matching efficiency, and energy loss efficiency. The background particles, knocked-out from material, are removed in isobaric collisions. However, this correction is not applied at 3 GeV due to the lacking of anti-protons needed for evaluating the correction factors. Since the feed-down contribution from the weak decay of strange baryons to protons is only about 1.5\% at 3 GeV, the feed-down correction to the proton yield is not applied.

\section{Results}

\subsection{Light nuclei production in Isobaric collisions at {\sNN} = 200 GeV}
\begin{figure}[htb]
\centering
	\includegraphics[width=12cm]{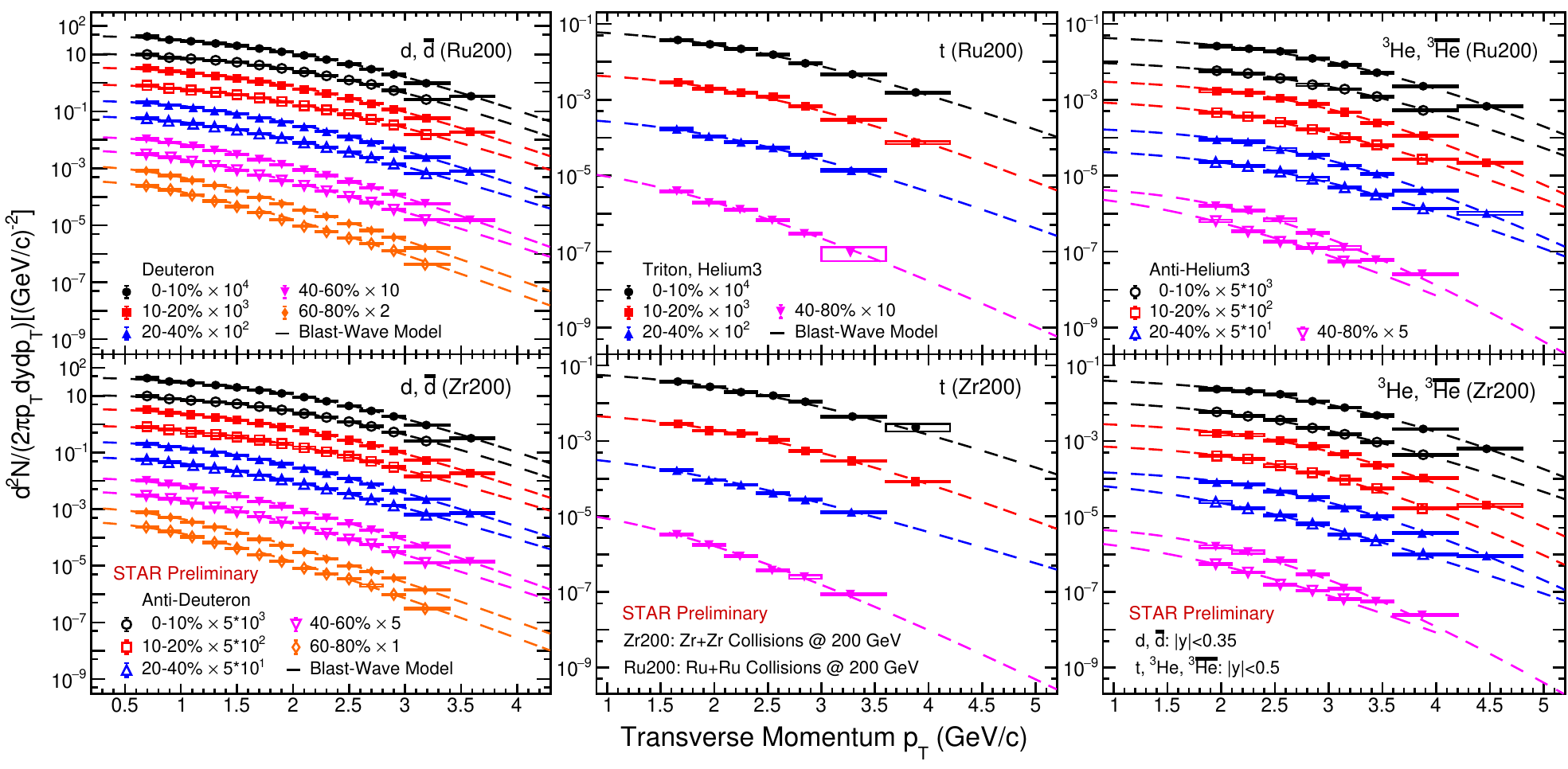}
\caption{\small Transverse momentum spectra of $d$, $\overline{d}$, $t$, $^{3}\mathrm{He}$, and $\overline{^{3}\mathrm{He}}$ measured at mid-rapidity in 0-10\%, 10-20\%, 20-40\%, 40-60\%, and 60-80\% (40-80\%) isobaric collisions at {\sNN} = 200 GeV. The dashed lines correspond to individual fits to the distributions with the Blast-Wave model function. The hollow boxes represent the systematic uncertainties.}
\label{fig:spectra200}
\end{figure}

The transverse momentum spectra of $d$, $\overline{d}$, $t$, $^{3}\mathrm{He}$, and $\overline{^{3}\mathrm{He}}$ at mid-rapidity in isobaric collisions at {\sNN} = 200 GeV in 0-10\%, 10-20\%, 20-40\%, 40-60\%, and 60-80\% (40-80\% for light nuclei with 3-nucleons) centrality bins are shown in Fig~\ref{fig:spectra200}. With very high statistics ($\sim$ 2 billion for each collision system), the statistical error is smaller than the marker size. In order to extrapolate the spectra to low and high $p_{T}$, the distributions are fit individually with the Blast-Wave model function~\cite{Schnedermann:1993ws}. 

\begin{figure}[htb]
\begin{minipage}[t]{0.52\linewidth}
\centering
	\includegraphics[width=5.8cm]{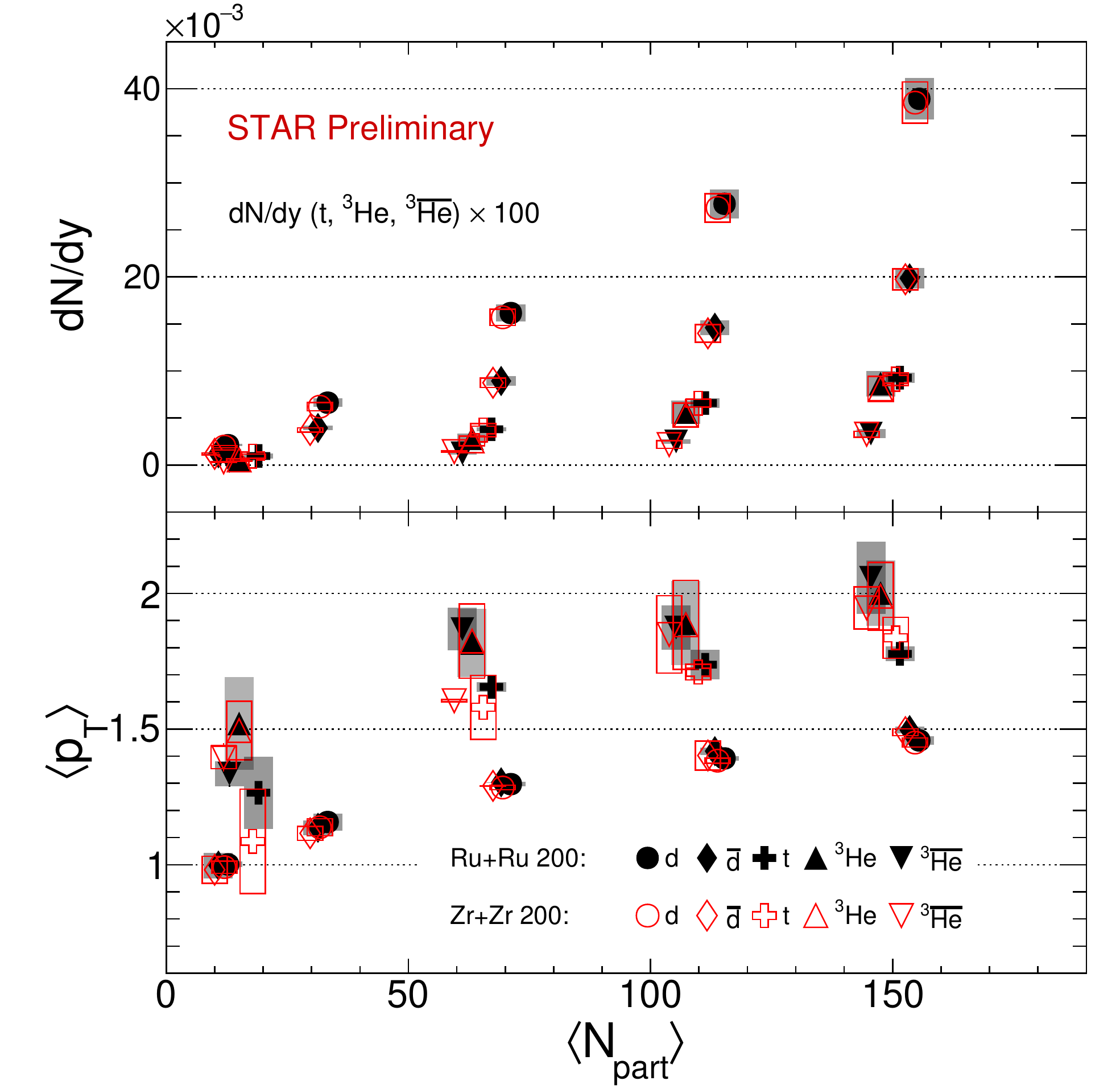}
\end{minipage}
\begin{minipage}[t]{0.05\linewidth}
\centering
	\includegraphics[width=5.62cm]{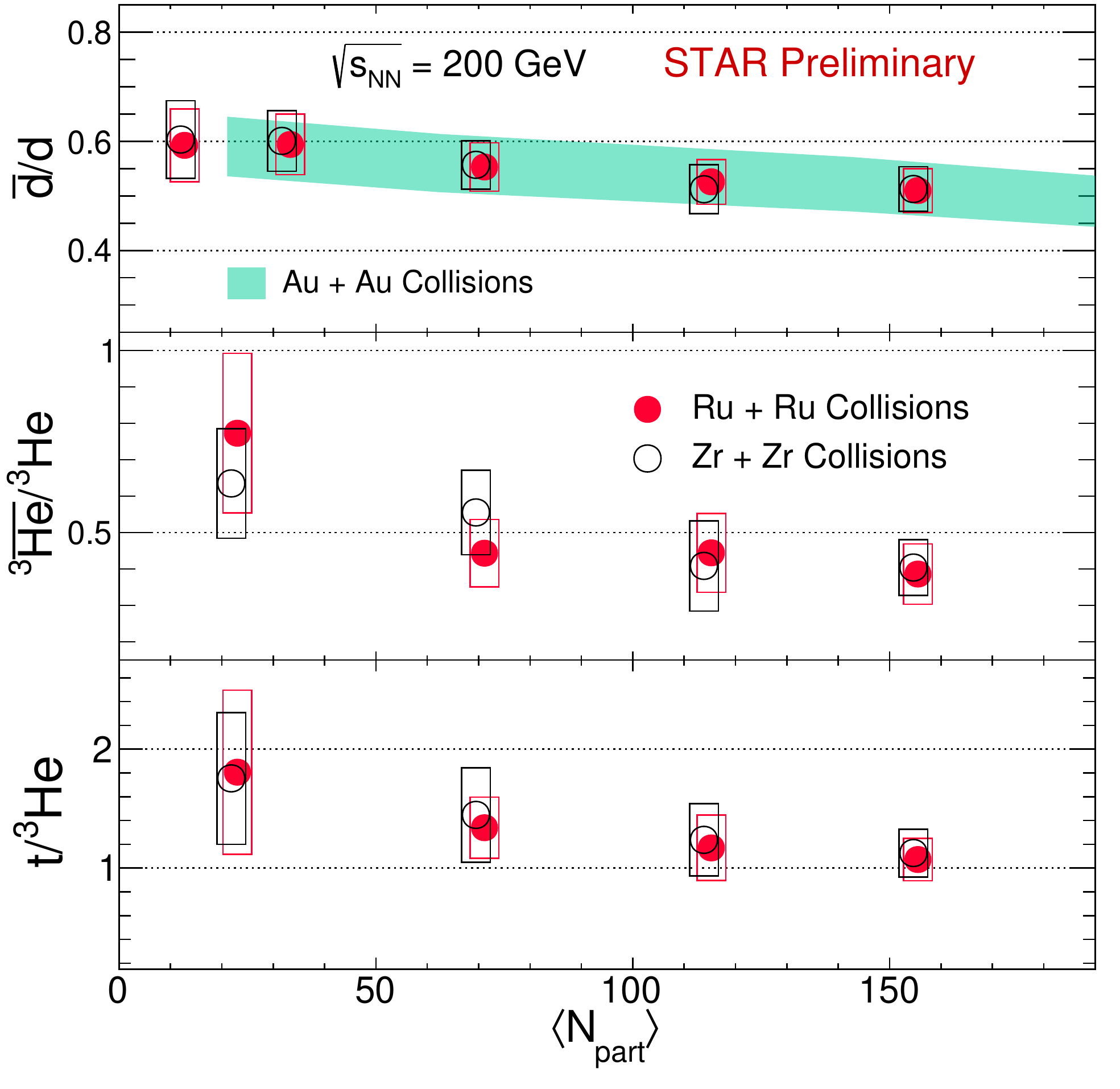}
\end{minipage}
\caption{\small Centrality dependence of dN/dy, $\langle p_{T}\rangle$, and particle ratios in isobaric collisions at {\sNN} = 200 GeV. The boxes and bands represent systematic uncertainties. The green band on the right plot is the $\overline{d} / d$ ratio in Au+Au collisions.}
\label{fig:dndy_Isobar}
\end{figure}

Figure~\ref{fig:dndy_Isobar}, left panel shows the rapidity density (dN/dy) and mean transverse momentum ($\langle p_{T}\rangle$) versus the average number of participating nucleons ($\langle N_{part}\rangle$) for $d$, $\overline{d}$, $t$, $^{3}\mathrm{He}$, and $\overline{^{3}\mathrm{He}}$. Both the dN/dy and $\langle p_{T}\rangle$ of each particle are consistent between Ru+Ru and Zr+Zr collisions. The particle ratios ($\overline{d} / d$, $\overline{^{3}\mathrm{He}} / ^{3}\mathrm{He}$, and $t / ^{3}\mathrm{He}$) are shown in the right panel. The $\overline{d} / d$ ratios in isobaric collisions are consistent with those in Au+Au collisions (green bands)~\cite{STAR:2008med} within uncertainties. All the ratios show an increasing trend from central to peripheral collisions.

\subsection{Light nuclei production in Au+Au collisions at {\sNN} = 3 GeV}
The Au+Au collisions at {\sNN} = 3 GeV with a fixed-target mode allows us to access the QCD phase structure at high baryon density regions with $\mu_{B} \sim$ 750 MeV, where the production of light nuclei is abundant.

\begin{figure}[htb]
\centering
	\includegraphics[width=12cm]{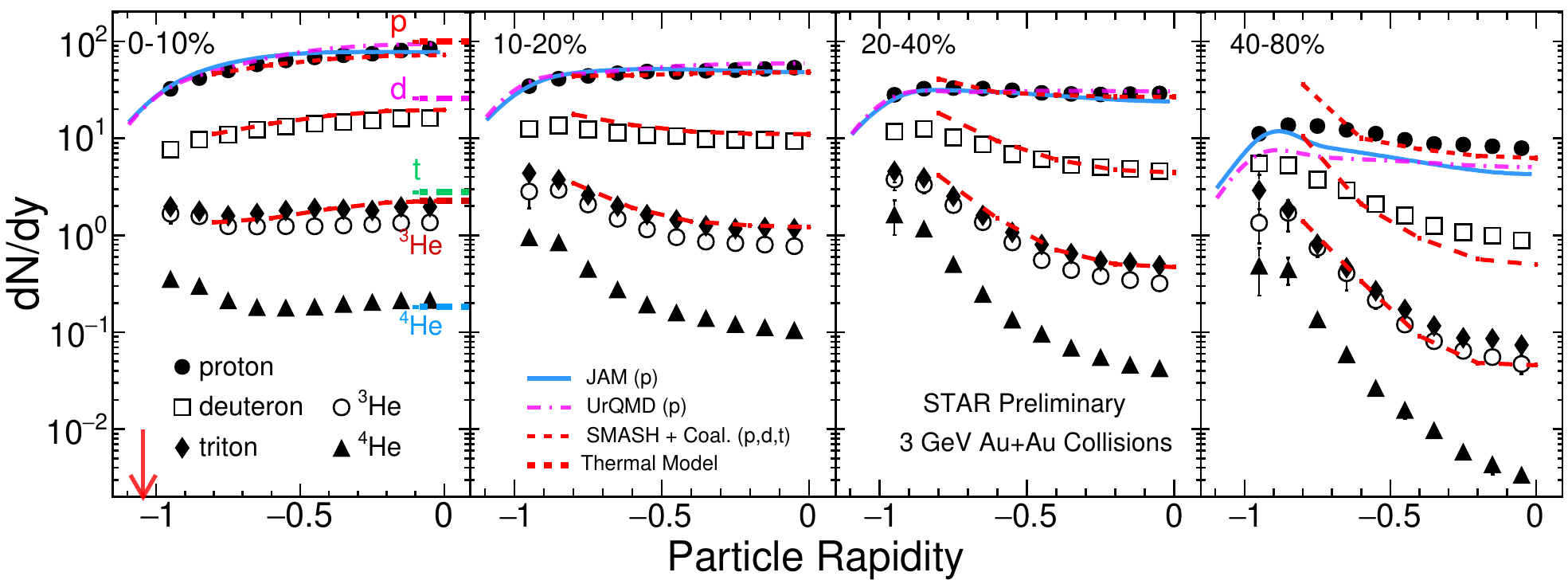}
\caption{\small The rapidity dependence of dN/dy for different centrality bins in Au+Au collisions at $\sqrt{s_{\mathrm{NN}}}$ = 3 GeV. The color lines are hadronic transport model (JAM, SMASH and UrQMD) calculations and thermal model results.}
\label{fig:dndy_3GeV}
\end{figure}

Figure~\ref{fig:dndy_3GeV} shows the dN/dy distributions of $p$, $d$, $t$, $^{3}\mathrm{He}$, and $^{4}\mathrm{He}$ in 0-10\%, 10-20\%, 20-40\%, and 40-80\% Au+Au collisions at $\sqrt{s_{\mathrm{NN}}}$ = 3 GeV. The target rapidity at this energy is at $y_{target} = -1.045$ (red arrow). For each particle, the dN/dy shows significant centrality and rapidity dependences. In more peripheral collision, dN/dy shows a peak near the target rapidity, which is caused by the interplay between produced nucleons and transported nucleons. As shown in colored lines, the hadronic transport models (JAM~\cite{Nara:2019crj}, SMASH~\cite{Weil:2016zrk}, and UrQMD~\cite{Bass:1998ca}) yield similar rapidity trends as experimental data except for the 40-80\% centrality bin. Calculating the formation probability by the Wigner function~\cite{Zhao:2021dka}, the rapidity density distributions of $d$ and $t$ are also well described by the SMASH model. In central collisions (0-10\%), we can extract thermal model parameters from measured hadron yields and consequently predict the light nuclei yields, which are shown as the short lines and seen to overestimate experimental data except for $^{4}\mathrm{He}$.

\begin{figure}[htb]
\centering
	\includegraphics[width=9.4cm]{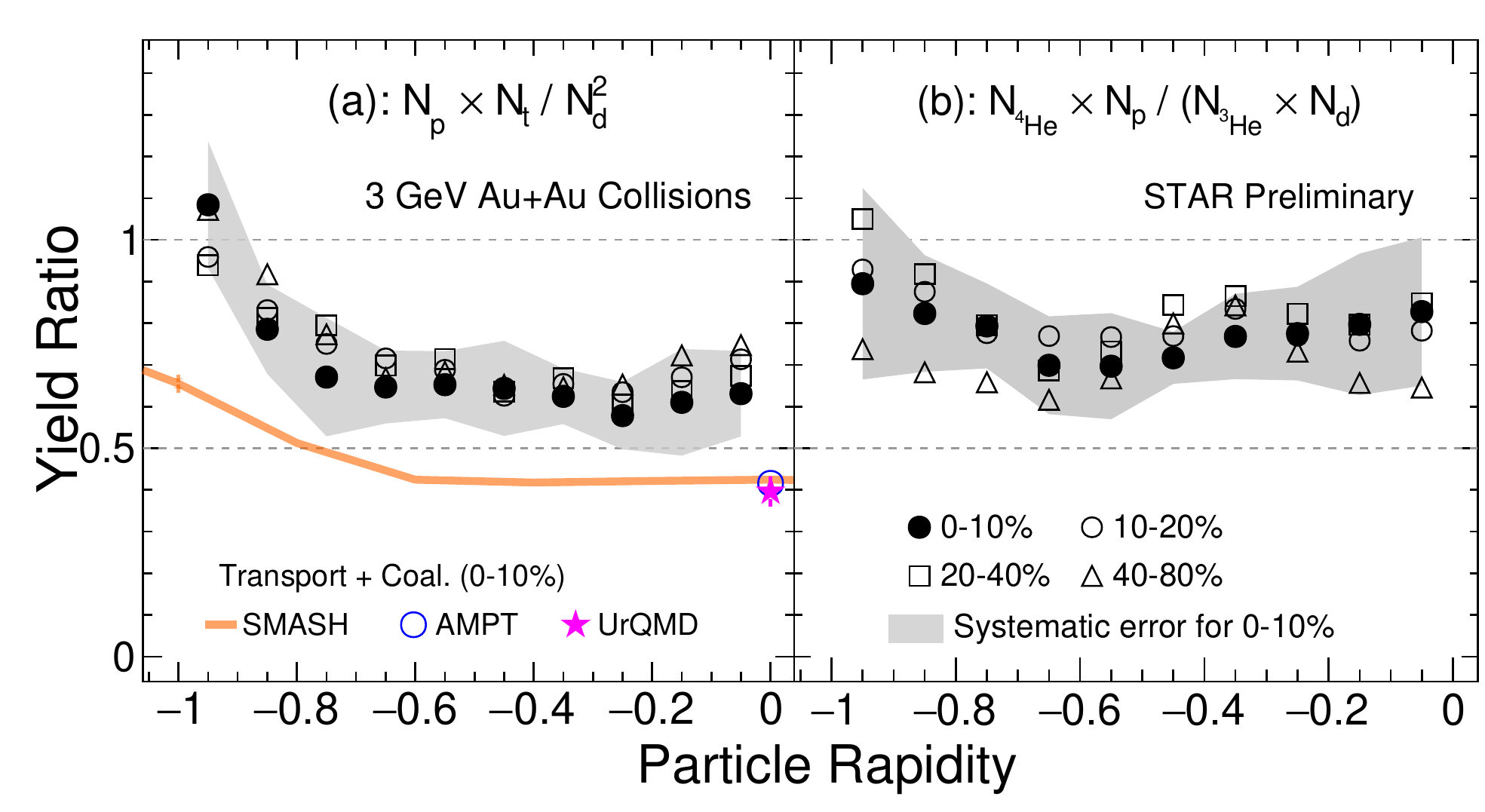}
	\caption{\small Rapidity dependence of the yield ratios in different centrality bins in $\sqrt{s_{\mathrm{NN}}}$ = 3 GeV Au+Au collisions. The gray bands represent the systematic uncertainty for 0-10\%. The $N_{p} \times N_{t} / N_{d}^{2}$ calculated by transport models (SMASH, AMPT, and UrQMD) are shown by colored markers.}
\label{fig:yield_ratio_cen}
\end{figure}

Figure~\ref{fig:yield_ratio_cen} shows the rapidity dependence of the yield ratios for $N_{p} \times N_{t} / N_{d}^{2}$ and $N_{^{4}\mathrm{He}} \times N_{p} /\left(N_{^{3}{\mathrm{He}}} \times N_{d}\right)$ in 0-10\%, 10-20\%, 20-40\%, and 40-80\% centralities, respectively. There is no obvious centrality dependence for each yield ratio. In contrast to the centrality trend, there is a clear increasing tendency towards target rapidity ($-1.0<y<-0.6$). The values of $N_{p} \times N_{t} / N_{d}^{2}$ calculated by transport models are lower than the experimental measurements, but the SMASH model gives a similar rapidity dependence.

\begin{figure}[htb]
\centering
	\includegraphics[width=6.7cm]{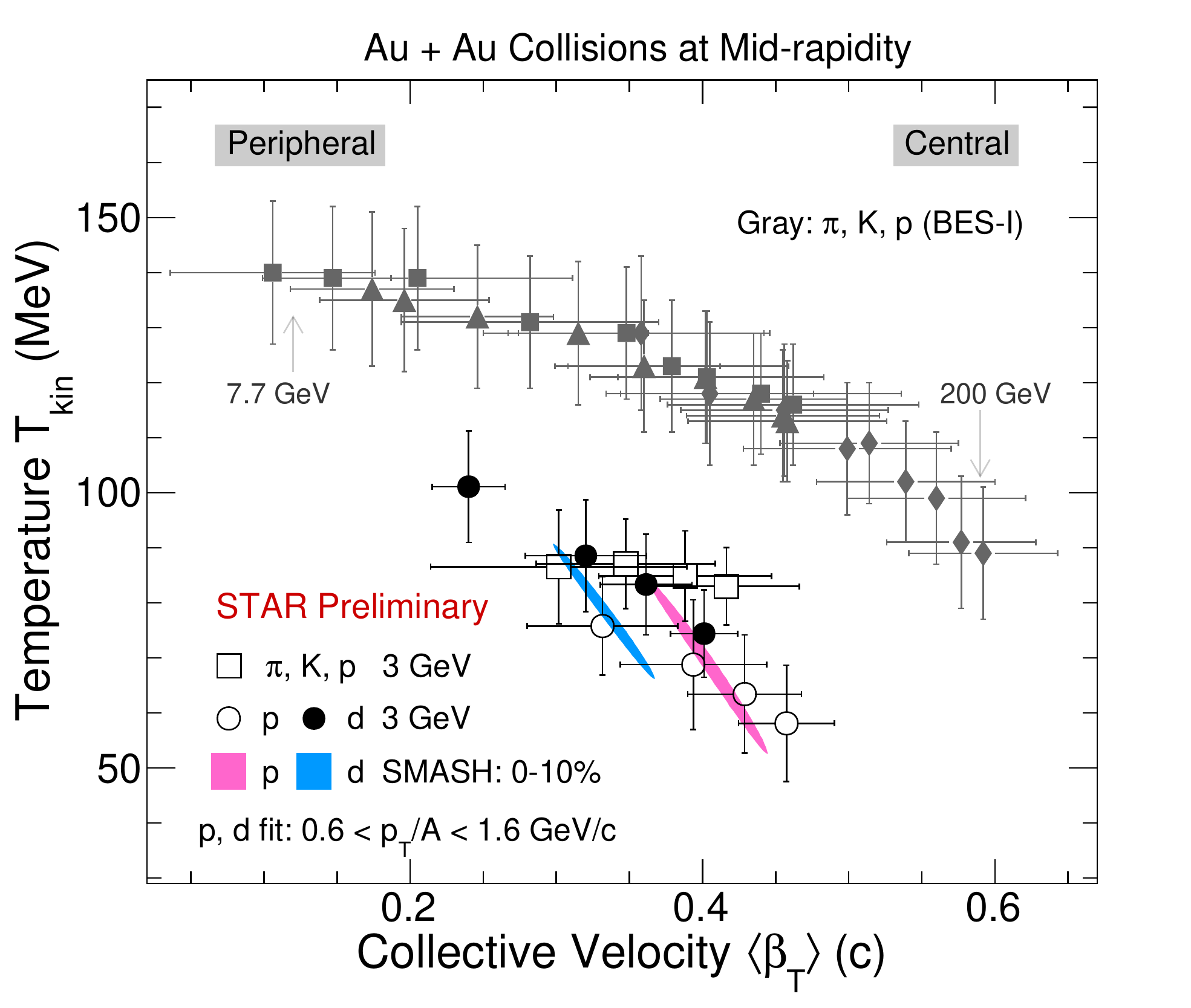}
\caption{\small Kinetic freeze-out parameters ($T_{\mathrm{kin}}$ and $\langle\beta_{T}\rangle$) dependence of particles at mid-rapidity for different centrality bins in Au+Au collisions at $\sqrt{s_{\mathrm{NN}}}$ = 3 GeV. The vertical lines represent the systematic uncertainties.}
\label{fig:Tbeta_all}
\end{figure}

Through fitting the $p_{T}$ spectra of particles by the Blast-Wave model~\cite{Schnedermann:1993ws}, we can extract the kinetic freeze-out parameters (temperature $T_{\mathrm{kin}}$, average radial flow velocity $\langle\beta_{T}\rangle$). In the current analysis, the Blast-Wave model is assumed to be the underlying boost-invariant longitudinal dynamics. 
Figure~\ref{fig:Tbeta_all} shows the $T_{\mathrm{kin}}$ versus $\langle\beta_{T}\rangle$ distribution of particles at mid-rapidity in different centrality bins. $T_{\mathrm{kin}}$ ($\langle\beta_{T}\rangle$) of the deuteron is systematically higher (lower) than that of the proton at 3 GeV as the black solid and open circles show. A similar trend is seen in the SMASH model calculation, shown as colored contours. Comparing the parameters of light hadrons ($\pi$, $K$, $p$) to the results from higher energies, as indicated by the gray markers, a different trend is seen at 3 GeV (open square), which seems to imply that the hot and dense medium created in 3 GeV collisions could be different from those at higher energy collisions.

\section{Summary}
In summary, we report the light nuclei ($d$, $\overline{d}$, $t$, $^{3}\mathrm{He}$, and $^{3}\overline{\mathrm{He}}$) production in isobaric (Ru+Ru, Zr+Zr) collisions at $\sqrt{s_{\mathrm{NN}}}$ = 200 GeV. It is observed that the yields of light nuclei are consistent between Ru+Ru and Zr+Zr collisions within uncertainties. Furthermore, we present the proton and light nuclei ($d$, $t$, $^{3}\mathrm{He}$, and $^{4}\mathrm{He}$) production in Au+Au collisions at $\sqrt{s_{\mathrm{NN}}}$ = 3 GeV. The dN/dy of those particles show strong centrality and rapidity dependences.  The compound yield ratios exhibit an increasing trend from middle to target rapidity, and the results of the transport models (AMPT, SMASH, and UrQMD) show lower values than the data.
Finally, the freeze-out parameters ($T_{kin}$, $\langle \beta_{T}\rangle$), extracted using the boost-invariant Blast-Wave model, show a different trend compared to the results from high energy. In addition, ($T_{kin}$, $\langle \beta_{T}\rangle$)  of deuterons are found to be (larger, smaller) than those of protons, indicating that there is no common freeze-out parameters among proton and deuteron. Hadronic transport model (SMASH) calculations reproduce the trend well. 

\section*{Acknowledgments}
We thank Dr. J. Aichelin, Dr. E. Bratkovskaya, Dr. J. Steinheimer and Dr. K. J. Sun for insightful discussions about production mechanism of light nuclei. This work was supported by the National Key Research and Develop- ment Program of China (2020YFE0202002 and 2018YFE0205201), the National Natural Science Foundation of China (12122505, 11890711) and the Fundamental Research Funds for the Central Universities(CCNU220N003).

\small
\bibliography{proceeding} 

\end{document}